\documentclass[prl,twocolumn,10pt,showpacs,letterpaper]{revtex4}
\usepackage{graphicx}
\begin{document}
\title[Strangelets as cosmic rays]{Strangelets as Cosmic Rays beyond the Greisen-Zatsepin-Kuzmin Cutoff}
\author{Jes Madsen and Jonas M\o ller Larsen}
\affiliation{Department of Physics and Astronomy, University of Aarhus, DK-8000 \AA rhus C, Denmark}
\pacs{12.38.Mh, 12.39.Ba, 24.85.+p, 98.70.Sa}

\begin{abstract}
Strangelets (stable lumps of quark matter) can have masses and charges much
higher than those of nuclei, but have very low charge-to-mass ratios. This is
confirmed in a relativistic Thomas-Fermi model. The high charge allows
astrophysical strangelet acceleration to energies orders of magnitude higher
than for protons. In addition, strangelets are much less susceptible to the
interactions with the cosmic microwave background that suppress the flux of
cosmic ray protons and nuclei above energies of $10^{19}$--$10^{20}$ eV (the
GZK-cutoff). This makes strangelets an interesting possibility for explaining
ultra-high energy cosmic rays.

\end{abstract}
\date{November 27, 2002}
\maketitle

A long standing puzzle in cosmic ray physics has been the nature of ultra-high
energy cosmic rays at energies well above $10^{20}$eV. Protons and nuclei that
are known to be responsible for a significant fraction of the cosmic ray flux
at lower energies cannot easily be accelerated to these energies, and if they
are, their interactions with photons in the cosmic microwave background
radiation are sufficiently energetic to lead to photo-pion and photo-pair
production, thereby reducing the energy. For nuclei photo-disintegration is an
additional important factor. This leads to an effective cutoff in flux
(originally suggested by Greisen, Zatsepin and Kuzmin \cite{GZK}; the
GZK-cutoff) at energies around $10^{19}$eV for protons, and $10^{20}$eV for
the heaviest, abundant stable nucleus, iron. Nevertheless, the observed flux
shows no clear cut at these energies (though the number of events is small,
and some inconsistency between different experiments exist), with observed
cosmic ray energies as high as $3\times10^{20}$eV \cite{crreview}. Many
suggestions have been made for the nature of cosmic rays beyond the
GZK-cutoff, ranging from very nearby sources (though there is no consensus
regarding anisotropy in the data at the very highest energies), to
\textquotedblleft new physics\textquotedblright\ such as decaying ultraheavy,
supersymmetric particles \cite{crreview}.

Here we suggest an alternative explanation for cosmic rays at the very highest
energies---strangelets (stable lumps of quark matter with roughly equal
numbers of up, down, and strange quarks), that due to a high mass and charge
but low charge-to-mass ratio in a natural way circumvent the acceleration
problem, and moves the GZK-cutoff to much higher energies \cite{APPB}.

The possibility that strange quark matter may be absolutely stable
\cite{bodmer71,madsenreview} has gained renewed attention recently.
Chandra X-ray observations of one unusually cold pulsar and
another unusually small one \cite{chandra} fit predictions for
strange stars made of absolutely stable strange quark matter, though several
other explanations are possible as well. An investigation of
earthquake data has found unusual seismic events consistent with
the passage of macroscopic quark matter lumps through Earth
\cite{seismic}. And predictions of increased quark
matter stability in a color-flavor locked phase \cite{alford01} has made
stable strange quark matter in bulk, and perhaps in smaller
strangelets \cite{madsen2001} more likely than previously believed.

The present investigation shows that strangelets have properties that in a
natural way circumvent the objections against protons and nuclei as ultra-high
energy cosmic rays. Strangelets, whether made of \textquotedblleft
ordinary\textquotedblright\ or color-flavor locked quark matter, have very low
charge-to-mass ratios, but their absolute charges and masses can be extremely
high compared to nuclei. Cosmic ray acceleration in astrophysical sources
(e.g. in shock waves) is expected to cutoff at some high particle rigidity
$R=p/Z$, where $R$, $p$ and $Z$ denote particle rigidity, momentum, and charge
\cite{crreview}. This rigidity is given by the value of $R$ where the particle
Larmor radius in the magnetic field (which is proportional to $R$) exceeds the
size of the accelerator. But for relativistic energies $p=E$, where $E$ is the
energy, so $E_{\mathrm{max}}=ZR_{\mathrm{max}}$. Thus, the higher $Z$, the
higher $E_{\mathrm{max}}$. For typical cosmic ray nuclei, $Z\leq26$, and
certainly $Z<100$. This makes it essentially impossible to reach the highest
cosmic ray energies observed by means of astrophysical source acceleration of
protons or nuclei, given the astrophysical source limitations on
$R_{\mathrm{max}}$ \cite{crreview}. As shown below, much higher charge
and therefore maximal acceleration energy can be achieved for strangelets.

For protons the GZK-cutoff is caused by photo-pion production. Even though the
average photon energy in the 3K cosmic microwave background radiation is as
low as $E_{\mathrm{3K}}\approx7\times10^{-4}$eV, an ultra-high energy cosmic
ray proton with a Lorentz-factor $\gamma_{\pi}>m_{\pi}/E_{\mathrm{3K}}%
\approx10^{11}$ will reach the threshold for the processes $p+\gamma
\rightarrow\pi+\mathrm{nucleon}$, leading to significant energy loss and an
expected drop in cosmic ray flux at proton energies above $\gamma_{\pi}%
m_{p}\approx10^{20}$eV (detailed calculations show that the drop actually sets
in at energies more than an order of magnitude smaller \cite{crreview}).

For nuclei the photo-pion energy loss cross-section goes up as $A^{2/3}$, but
the cutoff energy increases linearly with $A$. However, photo-disintegration
of nuclei has a threshold of only 10MeV rather than $m_{\pi}$ in the rest
frame of the nucleus, so the cosmic ray flux of nuclei should drop at a
Lorentz factor of order $10^{10}$, corresponding to $E_{\mathrm{GZK}}%
\approx10^{19}A$eV (again, detailed calculations result in a drop at energies
somewhat lower than found by these simple estimates).

Another important energy loss mechanism is photo-pair production, which is
possible when the microwave background photon in the cosmic ray restframe is
seen to have an energy in excess of $2m_{e}\approx1$MeV. The threshold Lorentz
factor for this process is $10^{9}$, and the energy for baryon number $A$ is
$E_{\mathrm{pp}}\approx10^{18}A$eV. The energy loss rate is proportional to
$Z^{2}A^{-1}\propto Z$ for nuclei with $Z\approx A/2$, so this mechanism is
significantly more important for nuclei like iron than for protons
\cite{crreview}.

Strangelets may exist in two possible varieties: \textquotedblleft
Ordinary\textquotedblright\ strangelets \cite{bodmer71,farhi84,madsenreview}%
\ and color-flavor locked strangelets \cite{madsen2001}. Both types of objects
consist of almost equal numbers of up, down, and strange quarks so that the
quark charges nearly cancel. The stability of strangelets depends on the
strong interaction binding (in phenomenological models characterized by
quantities like the bag constant and the one-gluon exchange coupling constant)
and the quark masses (especially the heaviest, strange quark). Finite size
effects (surface tension and curvature energy) generally decrease stability at
low baryon numbers relative to bulk strange quark matter, though increased
stability occurs near closed shells. \textquotedblleft
Ordinary\textquotedblright\ strangelets are stable for a restricted range of
parameters, and if quark matter is in a color-flavor locked state, as seems to
be the case at asymptotically high density \cite{alford01}, the stability is
improved by a significant binding energy in Cooper pairs coupling quarks of
different flavor and color quantum numbers.

\textquotedblleft Ordinary\textquotedblright\ strangelets have charge
$Z\approx0.1A$ for $A\ll150$ and $Z\approx8A^{1/3}$ for $A\gg150$
\cite{farhi84,heiselberg}, and color-flavor locked strangelets have
$Z\approx0.3A^{2/3}$ \cite{madsen2001}. In both cases the charge is much
smaller than that of nuclei of the same mass number, $A$, because strangelets
have almost equal numbers of up, down, and strange quarks so that the quark
charges nearly cancel. But strangelet masses can be very large (in principle
as large as the baryon number of a gravitationally unstable strange star,
$A_{\max}\approx2\times10^{57}$), so the quark charge $Z$ can reach values
much higher than those known for nuclei. What is here called the
\textquotedblleft quark charge\textquotedblright\ is the total charge of the
quark core of the strangelet. For color-flavor locked strangelets this charge
is contributed by quarks only, but \textquotedblleft
ordinary\textquotedblright\ strangelets with $A>10^{7}$ contain a small
electron fraction in weak equilibrium; this has not
been self-consistently included in the charge-mass relation.

Like nuclei may bind electrons to form neutral atoms, so a charge $Z$
strangelet may bind electrons and create a neutral atom-like system, but to be
accelerated as ultra-high energy cosmic rays, atoms or strangelets must be
ionized. Whereas atoms can be fully ionized to the nuclear charge, strangelet
core charges can be high enough that one has to worry about QED-effects. Of
relevance for the ultra-high energy cosmic rays is not the total quark charge
(as defined above), but rather the net screened charge when electrons from
QED-effects are taken into account. This screened charge is in practice the
net charge available for electromagnetic acceleration and interaction of a
relativistic cosmic ray. In quantum electrodynamics, the maximum unscreened
point charge is $1/\alpha\approx137$. For higher charge, electron-positron
pairs are created in the vacuum, the positron leaving the system, but the
electron remaining to screen the central positive charge. For extended systems
like nuclei, the maximum charge goes up in principle \cite{mr75}, but no
stable high charge nuclei seem to exist.

Strangelets have a lower charge density than nuclei, and color-flavor locked
strangelets have the charge located in a thin surface layer. This leads to the
expectation that the importance of charge screening due to electrons formed by
QED-effects is less pronounced than for nuclei \cite{farhi84}. This has been
confirmed by relativistic Thomas-Fermi model calculations (following a
procedure similar to the one adopted for nuclei in \cite{mr75}). For a given
unscreened charge above a few hundred, the net screened charge is highest for
\textquotedblleft ordinary\textquotedblright\ strangelets with the lowest
charge density, and lowest (i.e. most affected by screening) for nuclear
matter which has the highest charge density. The effects for color-flavor
locked strangelets are intermediate.

Results for the net unscreened and screened charge of nuclear matter and
strangelets are shown in Figure 1 as a function of baryon number, $A$.
Screened charges of several thousand are easily reached, and there is no
formal maximum charge, though the screened charge increases only slowly with
unscreened charge for high $Z$.

Relativistic strangelets are expected to be maximally ionized, that is, to a
net charge comparable to the one shown in Figure 1. In contrast
non-relativistic strangelets are expected to be charge neutral or have a small
net charge. This is the reason why we mainly discuss relativistic strangelets
in the present investigation. Non-relativistic strangelets could also have
kinetic energies in the range of interest (above $10^{20}$eV) if their masses
are sufficiently large. They would be harder to accelerate, but would avoid
the GZK-constraint by the same arguments discussed for relativistic
strangelets. We note that a strangelet moving in the galactic gravitational
potential with speed of order 300 km/sec would have kinetic energy in the
interesting range if its baryon number were of order $10^{17}$ (corresponding
to a mass of 0.1\ microgram).

Comparing with nuclei, the strangelet properties discussed above lead to the
following observations for relativistic strangelets:

1) For astrophysical acceleration mechanisms limited by a maximum rigidity
$R_{\mathrm{max}}$, strangelets can reach higher energies than nuclei, since
$E_{\mathrm{max}}=ZR_{\mathrm{max}}$.

2) The photo-pion and photo-disintegration energy cutoff moves upward
proportional to $A$, as $E\approx5\times10^{18}A$ eV for the latter. Thus
high-mass strangelets are not influenced by these processes at the energies of
interest.

3) Photo-pair production also sets in at energies that scale with $A$, and
furthermore, the energy loss rate is proportional to $Z^{2}A^{-1}$, which is
$\propto A\propto Z$ for nuclei, $\propto A^{1/3}\propto Z^{1/2}$ for
color-flavor locked strangelets, and $\propto A^{-1/3}\propto Z^{-1}$ for
``ordinary'' strangelets, in all cases using the unscreened charge-mass relations,
which significantly overestimate the relevant net charge of the system, c.f.
Figure 1. Therefore, strangelets are much less susceptible to photo-pair
energy loss than are nuclei. 
%The possibility that a $Z^2A^{-1}$ scaling of
%the energy loss rate could allow ``ordinary'' strangelets to have energies
%above the GZK-cutoff has also been pointed out in \cite{APPB}.

4) At a given energy cosmic ray strangelets will be more isotropically
distributed than protons and nuclei, since the radius of gyration (Larmor
radius) in the galactic magnetic field ($B\approx3\mu G$) is $r_{\mathrm{gyro}%
}=36\mathrm{kpc}\left(  E/10^{20}\mathrm{eV}\right)  \left(  3\mu G/B\right)
Z^{-1}$ compared to a galactic radius of $r_{\mathrm{gal}}\approx10$ kpc.
Thus, low-$Z$ nuclei and protons have $r_{\mathrm{gyro}}\approx
\ r_{\mathrm{gal}}$, whereas higher-$Z$ strangelets have $r_{\mathrm{gyro}%
}<\ r_{\mathrm{gal}}$ and therefore appear to be more isotropically
distributed (arrival directions do not point back to the source) for the
energies of interest here, unless they originate from local sources. There is
presently some discussion as to whether or not there is evidence for
anisotropy in cosmic ray data at the highest energies \cite{crreview}, but
this could be a distinctive feature for the strangelet scenario when higher
statistics data emerge in the coming years.

5) Extended air shower signatures \cite{crreview} are consistent with
primaries of ultra-high energy cosmic rays being protons or ordinary nuclei.
In contrast, more massive hadronic objects like dust grains (which could have
a high net charge) are inconsistent with air shower data if they contain more
than a few thousand (probably even fewer) nucleons \cite{dust}. It is
therefore legitimate to wonder, why strangelets may avoid such limits. A
detailed study of this would require simulations of strangelet cosmic ray air
showers, but unfortunately a lot of the necessary interaction input physics is
at best poorly known since strangelet studies have only been performed within
simple phenomenological models. But estimates show, that strangelets differ
sufficiently from nuclei and dust to be viable candidates. Nuclei and dust
grains colliding with atmospheric oxygen or nitrogen will rapidly fragment
into their constituent nucleons. Each of these has roughly the original
Lorentz-factor, and therefore a kinetic energy lower by a factor $A$, so the
shower starts higher in the atmosphere (because of a larger geometrical cross
section of the primary) and develops more rapidly (because the particles have
less energy to deposit) than for a proton of the same total kinetic energy. A
strangelet may also start interacting high in the atmosphere, but the
geometrical cross section may be a factor of a few smaller than expected for
nuclear matter with similar $A$ because the density of strange quark matter
exceeds that of nuclear matter. This would cause mass $A$ strangelets to
interact like nuclei of significantly lower $A$, and therefore extend the
range of compatibility with air shower data. Furthermore, strangelets may to
some extent absorb and convert atmospheric nuclei into quark matter, and if
and when they fragment, some of the fragments may themselves be strangelets
with baryon numbers consistent with air shower data \cite{banerjee}. We
consider it possible, that the upper limit on $A$ for low-mass strangelets
consistent with the data could well be $10^{4}$ or higher, but it depends on
the input physics. Another window opens at higher $A$. Fragmentation of
strangelets requires the total energy added in inelastic collisions with
atmospheric nuclei to be at least of order the binding energy, which can be
some tens of MeV per nucleon. The total column density of matter required for
this \cite{column} is 
$x\approx4\times10^{-12}E_{B10}E_{20}^{-1}(\rho_{s}/\rho_{n})^{2/3}A^{4/3}$, 
where $x$ is measured in grams per cm$^{2}$, $E_{B10}$ is the
binding energy per baryon in units of 10 MeV, $E_{20}$ is the cosmic ray
kinetic energy in units of $10^{20}$ eV, and the factor $(\rho_{s}/\rho
_{n})^{2/3}$ takes into account that the strangelet cross section is smaller
than that of nuclei because the strangelet density exceeds that of nuclei.
Notice that strangelet cosmic rays will fragment easily for low $A$, but for
$A>10^{8-9}$ the critical column density becomes comparable to the
fragmentation depth normally expected for nuclei, and in fact strangelets with
$A\gtrsim10^{11}$, that are only mildly relativistic at these energies, could
penetrate to the surface without fragmentation ($x_{\mathrm{surface}}%
\approx10^{3}$ g/cm$^{2}$). Furthermore, even if strangelets fragment, the
shower development will be very different for fragmentation into $A$ nucleons
compared to fragmentation into low-mass strangelets. Heavier, non-relativistic
strangelets would be able to reach the surface, and the detailed development
of air showers would depend in a complicated manner on the electrostatic
interactions of the strangelet plowing its way through the atmosphere in
combination with the interactions of those hadrons and smaller strangelets,
that would be formed in inelastic collisions. At the present level of
understanding we conclude that strangelets with mass up to or somewhat
exceeding the limit for dust (maybe $10^{4}$ baryons), and with $A>10^{8-9}$
could be consistent with air shower observations, but that intermediate masses
may also be consistent if fragmentation preferentially leads to formation of
low-mass strangelets rather than nucleons.

Needless to say, many details that are beyond the scope of this investigation
must be considered to decide if strangelets really are the ultra-high energy
cosmic rays beyond the GZK-cutoff. Strangelet physics has so far only been
studied in phenomenological models like the MIT bag model, so the
understanding of strangelet stability, the strangelet charge-mass
relationship, strangelet disintegration etc, are all at a very crude level.
However, the general features of importance for the present investigation,
namely the possibility of high strangelet mass and charge, but low
charge-to-mass ratio seem robust. Details of the origin and propagation of
cosmic ray strangelets need further study \cite{madsen2002}, and the signature
of ultra-high energy strangelet air showers should be simulated and compared
to the observations of cosmic ray air showers at the highest energies, though
such simulations will be marred by the uncertainties in the underlying
strangelet physics. Sources of cosmic strangelets could be strange star
collisions in binary systems, supernova explosions, or perhaps gamma ray
bursts, and acceleration could take place in these and other suggested cosmic
ray engines, but again, many details remain to be studied.

But in spite of these issues we conclude based on the present understanding of
strangelet physics, that strangelets are interesting candidates for ultra-high
energy cosmic rays beyond the GZK-cutoff. They have properties which
circumvent both the acceleration problem, and the energy-loss problems facing
more mundane candidates like protons and nuclei. A crucial test of the
suggestion would be a direct measurement of the charge and/or mass of
ultra-high energy cosmic ray primaries. A search for cosmic ray strangelets at
much lower energies will take place with the AMS-02 experiment on the
International Space Station starting in 2005.

This work was supported by the Theoretical Astrophysics Center under
the Danish National Research Foundation, and by the Danish Natural Science
Research Council.

\begin{figure}
[b]
\begin{center}
\includegraphics[
height=3.9574in,
width=3.3938in
]%
{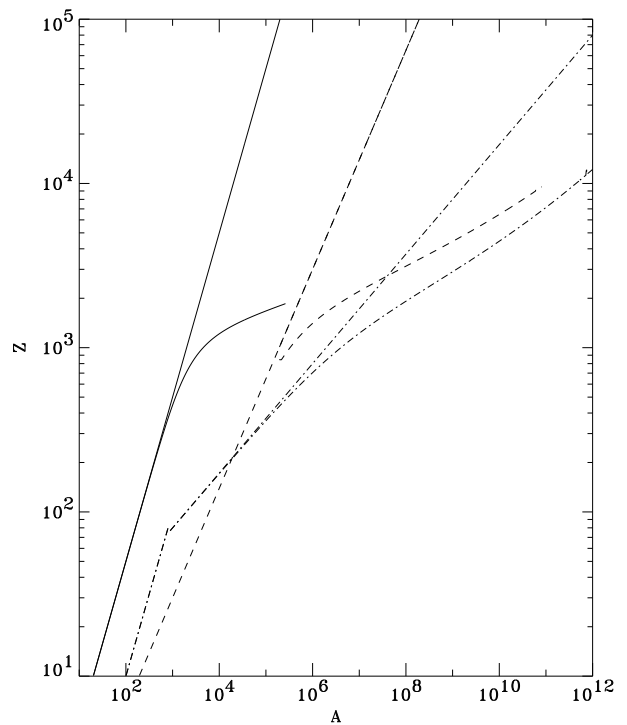}%
\caption{Charge as a function of mass for nuclear matter (full curves),
color-flavor locked strangelets (dashed curves), and ``ordinary'' strangelets
(dash-dotted curves). In each case the upper curve shows the unscreened
charge, whereas the lower curve is the net
screened charge from relativistic Thomas-Fermi calculations
including the supercritical vacuum.}%
\end{center}
\end{figure}


\begin{thebibliography}{99}
\bibitem{column}A strangelet with geometrical cross section 
$\sigma\approx \pi (1.1 {\rm fm} A^{1/3})^2 (\rho_n/\rho_s)^{2/3}$, where
$\rho_n$ and $\rho_s$ are the densities of nuclei and strange matter,
moving through a column density $x$ of 
atmospheric nuclei with baryon number $A_n$, hits 
$x\sigma/(A_n m_u)$ nuclei ($m_u$ is the unit mass). Each collision adds
a rest frame energy $E A_n/A$ to the strangelet, where $E$ is the strangelet
kinetic energy. Fragmentation requires a total energy
absorption $AE_B$, where $E_B$ is the binding energy per baryon,
or $x>A^2E_Bm_u/(\sigma E)$.
\end{thebibliography}
\end{document}